\documentclass{article}
\usepackage{spconf,graphicx,hyperref}
\usepackage{algorithm}
\usepackage{tabularx}
\usepackage{booktabs}
\usepackage{multirow} 
\usepackage{amsmath}
\usepackage{makecell}
\usepackage{algpseudocode}
\usepackage{amssymb}
\usepackage{cite}
\usepackage{enumitem}

\title{MiTa: A Hierarchical Multi-Agent Collaboration Framework with  Memory-integrated and Task allocation}

\name{XiaoJie Zhang$^{\star,\dagger}$, JianHan Wu$^{\dagger}$, Xiaoyang Qu$^{\dagger}$, Jianzong Wang$^{\dagger, \ddagger}$\thanks{This work is supported by Shenzhen-Hong Kong Joint Funding Project (Category A) under grant No. SGDX20240115103359001.\\ 
\indent$\hspace{1em}^{\ddagger}$Corresponding author is Jianzong Wang (jzwang@188.com)}}
\address{$^{\star}$Tsinghua Shenzhen International Graduate School, Tsinghua University, Shenzhen, China\\
        $^{\dagger}$Ping An Technology (Shenzhen) Co., Ltd., Shenzhen, China}

\begin{document}
%
\maketitle
\begin{abstract}

Recent advances in large language models (LLMs) have substantially accelerated the development of embodied agents. LLM-based multi-agent systems mitigate the inefficiency of single agents in complex tasks. However, they still suffer from issues such as memory inconsistency and agent behavioral conflicts.
To address these challenges, we propose MiTa, a hierarchical Memory-integrated Task allocative framework to enhance collaborative efficiency.
MiTa organizes agents into a manager–member hierarchy, where the manager incorporates additional allocation and summary modules that enable (1) global task allocation and (2) episodic memory integration. 
The allocation module enables the manager to allocate tasks from a global perspective, thereby avoiding potential inter-agent conflicts.
The summary module, triggered by task progress updates, performs episodic memory integration by condensing recent collaboration history into a concise summary that preserves long-horizon context.
By combining task allocation with episodic memory, MiTa attains a clearer understanding of the task and facilitates globally consistent task distribution. Experimental results confirm that MiTa achieves superior efficiency and adaptability in complex multi-agent cooperation over strong baseline methods.

\end{abstract}
\vspace{-0.05cm}
\begin{keywords}
Large Language Models, Embodied, Multi-Agent, Memory-integrated, Task allocation
\end{keywords}

\section{Introduction}
\vspace{-0.1cm}
\label{sec:intro}
The pursuit of embodied agents capable of autonomously accomplishing general tasks has long been a central objective in AI research\cite{embodied_machinery,embodied_research}. Recent advances in large language models (LLMs)\cite{gpt4o,qwen3} demonstrate strong abilities in language understanding\cite{language,iccv}, logical reasoning\cite{reason_embodied}, and task planning\cite{taskplan}, making them powerful drivers for embodied agents\cite{LLM-driven,acl} that integrate reasoning with perception and action.
Language-Planner\cite{zeroshot} and LLM-Planner\cite{llm-planner} explore high-level planning in embodied scenarios under zero-shot and few-shot settings. Octopus\cite{octopus} introduces a vision-language programmer linking high-level plans with real-world manipulation.
However, constrained by individual agents’ limited capabilities, these frameworks underperform in complex scenarios.

Inspired by human collaboration\cite{human_cooperation}, recent work studies coordination in embodied multi-agent systems\cite{survey_multi-agents}, leveraging communication and joint decision-making to outperform single agents in complex tasks. 
CoELA\cite{COELA} introduces a modular LLM-based framework for multi-agent cooperation. ProAgent\cite{proagent} enables infer teammates’ intentions and update its own beliefs. 
LGC-MARL\cite{icra} combines LLM planning and graph-based MARL to improve coordination efficiency and scalability in multi-agent systems.
However, these methods rely on limited memory and lack a centralized manager\cite{four_strusct}, hindering long-horizon coordination and causing information loss, poor context tracking, and task conflicts that lead to failures in complex long-horizon tasks.

\begin{figure*}[t]
\centering
\includegraphics[width=1\textwidth, trim=0.2cm 0cm 0.4cm 0.3cm, clip]{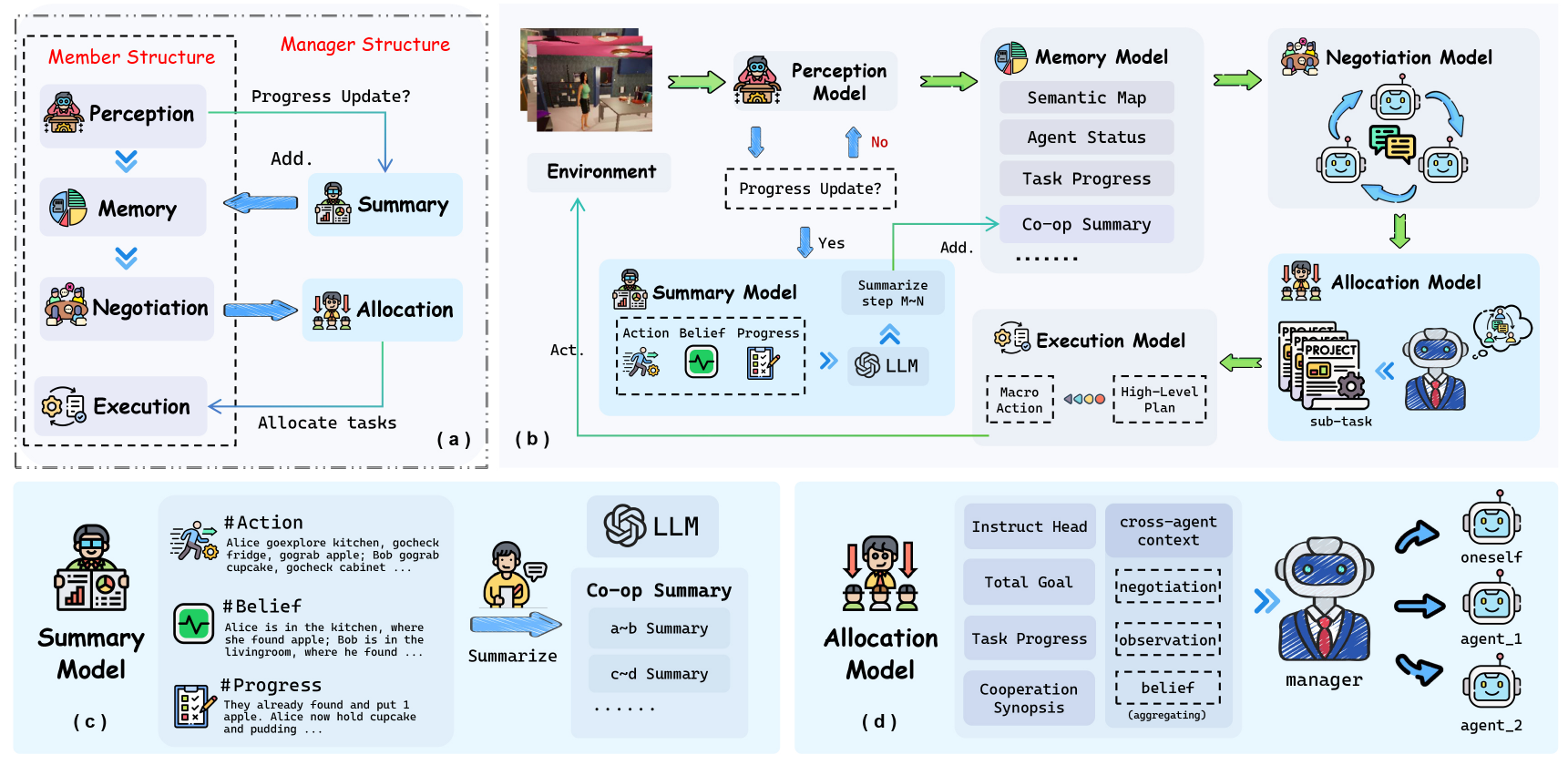} 
\vspace{-0.9cm}
\caption{ \textbf{Overview of the MiTa framework.} MiTa organizes agents into two roles: manager and member. (a) The overall workflow of MiTa, including the structures of both managers and members. (b) Detailed workflow of the manager agent. (c) Details of the Summary modules. (d) Structure of the prompt used in the allocation module.}
\label{main_fig}
\vspace{-0.3cm}
\end{figure*}

To address these limitations, we propose MiTa, a memory-integrated task allocation framework for multi-agent collaboration. MiTa adopts a hierarchical architecture with a centralized manager and multiple members. The manager incorporates two specialized modules, Allocation and Summary, to improve coordination efficiency and contextual consistency. The Allocation module leverages cross-agent context and recent collaboration updates to assign subtasks from a global perspective, reducing potential conflicts. To maintain long-horizon context, the Summary module distills collaboration updates and task progress into concise episodic memories upon state changes, which are stored persistently to guide future planning and avoid information loss caused by truncated context windows.
These designs leverage the strengths of LLMs in logical reasoning \cite{reason_embodied} and textual summarization \cite{summarization}, enhancing the flexibility and adaptability of multi-agent systems in complex environments.

\begin{itemize}[parsep=0.5pt, partopsep=0pt]
\item We propose MiTa, an LLM-guided hierarchical multi-agent framework, combining task allocation and memory integration for improved coordination.
\item We divide agents into a manager and members, where the manager leverages dedicated allocation and summary modules to orchestrate the team in a more structured and coordinated manner to accomplish tasks.
\item Experimental results show that MiTa outperforms state-of-the-art methods and can achieve near-optimal performance even when members use weaker models.
\end{itemize}

\vspace{-0.4cm}
\section{Task Formulation}
\label{sec:pagestyle}

Our embodied multi-agent cooperation task can be formulated as a Multi-agent 
Partially Observable Markov Decision Process (MPOMDP)\cite{MPOMDP}, which is defined as:
\vspace{-0.1cm}
\begin{equation} 
\mathcal{M}=(\mathcal{I},\mathcal{S},s_{0},\{\langle{O}_{i},{A}_{i}\rangle\},\mathcal{T},\Omega,R,\gamma)
\end{equation}

where $\mathcal{I}$ is a set of agents; $\mathcal{S}$ is a finite set of states with initial state distribution $s_{0}$; $\langle{O}_{i},{A}_{i}\rangle$ is the observation and action space of each agent; $\mathcal{T}$ is the transition probability function;  $\Omega(o\mid s^{\prime},a)$ is observation function; $R$ is the globally shared reward function; $\gamma$ is the discount factor.


In the MiTa framework, agents set $\mathcal{I} = \{1, \dots\ ,n\}$ operates in a shared state space $S$. Each agent $i$ receives a local observation $o_j \in O_i$, which together form a joint observation $\mathbf{o}\in O$ and update the team belief $b(s)$. Based on this belief, a centralized manager assigns a joint action $\mathbf{a}=\{a_1,\dots,a_n\}\in A$ from the agents’ respective action spaces. The environment then transitions according to $T(s' \mid s,\mathbf{a})$, produces a reward $R(s,\mathbf{a})$, and yields a new joint observation $\mathbf{o}'$.
This process iterates with discount factor $\gamma$, optimizing a joint policy $\pi(b)$ to maximize the expected long-term return.

\vspace{-0.1cm}
\section{method}
\vspace{-0.1cm}

\subsection{Overall Framework}
\vspace{-0.2cm}
Our proposed MiTa framework (Fig.\ref{main_fig}) categorizes agents into a manager and members. Each members include four modules: Perception, Memory, Negotiation, and Execution, while the manager adds Allocation and Summary modules to provide global guidance and maintain coordinated, goal-directed execution. 
During the task, Perception encodes environmental states into natural language and stores them in Memory.
As the task progresses, the manager leverages an LLM to summarize recent updates into a concise collaborative summary. Then Negotiation enables agents to propose candidate actions from their local perspectives, after which the manager allocates tasks from a global perspective. Finally, the Execution Module translates the high-level plan into executable macro actions. 
Further details of the allocation and summary modules are provided in Sec.\ref{method3.1} and Sec.\ref{method3.2}.


\vspace{-0.4cm}
\subsection{Negotiation-Aware Centralized Task Allocation}
\vspace{-0.1cm}
\label{method3.1}
Decentralized collaboration in multi-agent systems often causes redundancy, contention, and coordination failures, especially in complex multi-step tasks requiring agents' collaboration. To mitigate these challenges, we propose a structured collaboration mechanism that combines bottom-up negotiation with a top-down allocation strategy.

During the negotiation phase, each agent leverages LLM to generate a proposal $m_i^t$ regarding its next-step action, conditioned on its state, perception, memory, and task context. This process can be formally expressed as:

\begin{equation}
m_i^t=f_{LLM}\left(o_i^t,\mathcal{H}_i,p^{t},T\right)
\end{equation}
where $o$ denote the agent’s partial observation; $\mathcal{H}$ refers to the agent’s recent dialogue and action history; $p$ indicates the task progress; $T$ represents the global task objective.

Following negotiation, the allocation stage proceeds in two steps. We first assemble the cross-agent context $\mathcal{X}$ by collecting each agent’s proposal together with its local state and observation:


\begin{equation} 
\mathcal{X}^t=\sum_{i\in\mathcal{I}}{\{m_i^t,b_i^t,o_i^t\}}
\end{equation}
where $m$ denotes the proposal of each agent; $b$ and $o$ represents the agent’s belief and partial observation.

Given this context, the manager selects a globally coherent joint action by maximizing an LLM-instantiated scoring function over the joint action space:
\begin{equation} 
\label{eq.4}
\mathbf{a}^{t*}=\arg\max_{\mathbf{a}^{t}\in\mathbb{A}^d}\mathcal{M}\left(\mathcal{X}^t,\mathcal{C},p^{t},T\right)
\end{equation}
where $\mathcal{X}$ is the cross-agent context; $\mathcal{C}$ donates the collaborative summary; $\mathcal{M}$ is global operator mapping based on LLM.



\vspace{-0.2cm}
\subsection{LLM-Driven Episodic Memory Integration}
\vspace{-0.1cm}
\label{method3.2}
In multi-agent collaboration, agents typically require multiple rounds of interaction to accomplish complex long-horizon tasks. Each agent maintains only a partial observation $o_i^t$, and the team belief state $b_t$ must be inferred from the joint history $H_t=\{b_{0:t-1},a_{0:t-1},o_{0:t},\}$. However, prior frameworks typically approximate $H_t$ using a truncated window of recent interactions\cite{COELA,proagent}, which discards potentially relevant long-horizon dependencies, resulting in information loss and poor contextual tracking. 

\begin{figure}[htbp]
    \centering
    \vspace{-0.3cm}
    \includegraphics[width=0.49\textwidth]{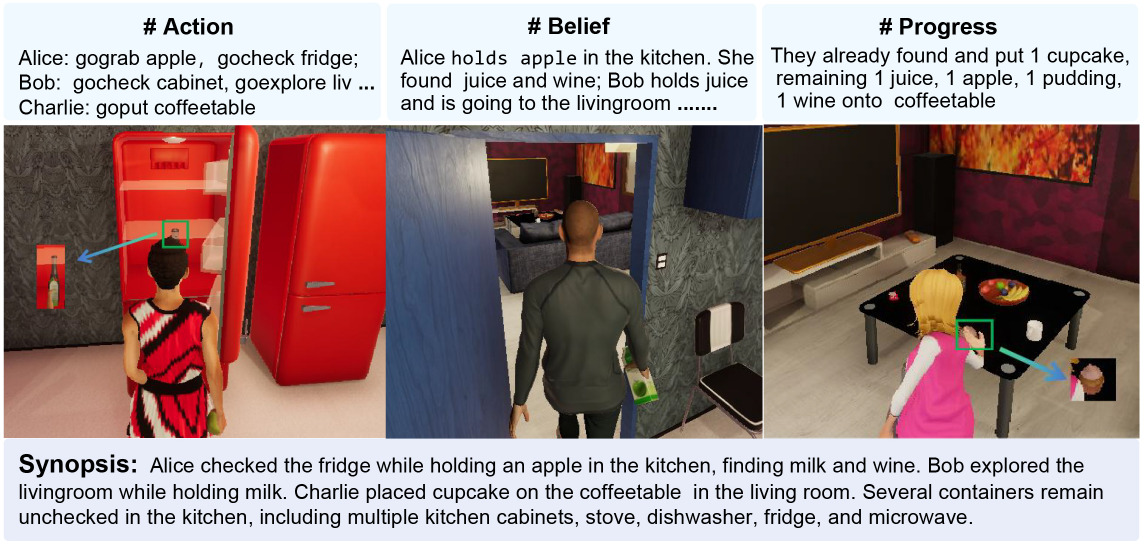} 
    \vspace{-0.8cm}
    \caption{Example of generating a collaborative summary}
    \vspace{-0.4cm}
    \label{syn}
\end{figure}

To address this limitation, we design a summary module that leverages the summarization\cite{summarization} capabilities of LLMs to integrate long-term episodic memory, as illustrated in Fig.\ref{syn}. Specifically, when a task progress update occurs, the module retrieves the joint action $\mathbf{a}$ and team belief state $b(s)$ accumulated since the last update from the memory module. These records, together with the task progress, are then composed into a structured prompt to generate a concise collaborative summary $\mathcal{C}$. The pseudocode is shown in Algorithm \ref{alg:progress_summary}. 

\begin{algorithm}[t]
\caption{Summary-based episodic memory integration}
\label{alg:progress_summary}
\begin{algorithmic}[1]  

\State \textbf{Input:} Agent set $\mathcal{I} = \{1, \dots, i\}$; Initial belief $b^0$; Initial state $s^0$; Initial task progress $p_0$; Total task $T$;
\State \textbf{Output:} Collaborative summary $\mathcal{C} = \{c_1, \dots\ ,c_n\}$;
\State Initialize: $t \gets 0$, $n \gets 0$, $p_{\text{last}} \gets p_0$, $\mathcal{S} \gets \{s^0\}$, $\mathcal{C} \gets \emptyset$;

\For{each episode}
    \State Update $\mathbf{a}^t = \{a_1^t,\dots,a_i^t\}$ using Eq.(\ref{eq.4})
    \State Environment transitions: $s^{t+1} \gets \text{Trans.}(s^t, \mathbf{a}^t)$
    \State New joint observation: $\mathbf{o}^{t+1} = \{o_1^{t+1},\dots,o_i^{t+1}\}$
    \State Belief update $b^{t+1}\gets$Belief$(b^t, \mathbf{a}^t,\mathbf{o}^{t+1}) $ 
    \State Evaluate current progress: $p_t \gets \text{Eval.}(b^{t+1},T)$
    
    \If{$p_t \neq p_{\text{last}}$}
        \State Retrieve history since the last summary:
        \State $\mathcal{H}_t = \{(a_i^\tau, b^\tau) \mid \tau \in (t_{\text{last}}, t], i \in \mathcal{I} \}$
        \State Record progress changes: $\Delta p \gets p_t - p_{\text{last}}$
        \State Generate summary: $c_n \gets \text{LLM}(\mathcal{H}_t, \Delta p)$
        \State Add $\{c_n\}$ into $\mathcal{C}$:  $\mathcal{C} \gets \mathcal{C} \cup \{c_n\}$
        \State Update: $p_{\text{last}} \gets p_t$, $t_{\text{last}} \gets t$,$n \gets n + 1$
    \EndIf
    \State $t \gets t + 1$
\EndFor
\State \textbf{return} $\mathcal{C}$
\end{algorithmic}

\end{algorithm}
\vspace{-0.2cm}

\begin{table*}[t]
  \centering
  \small
  \caption{Performance comparison of different methods across tasks with varying agent numbers under symbolic observation using GPT-4o. We report the Average Steps($\downarrow$) and Efficiency Improvement($\uparrow$), such as  34.4 (+68\%) on the C-WAH task.}
  \vspace{0.1cm}
  \label{tab 1}
    \begin{tabularx}{\textwidth}{@{}c|c|*{5}{>{\centering\arraybackslash}X}|c@{}}
      \toprule[1.2pt]
      \textbf{Num} & \textbf{Method} & \textbf{Wash Dishes} & \textbf{Put Groceries} & \textbf{Prepare a Meal} & \textbf{Set up Table} & \textbf{Prepare Tea} & \textbf{Average} \\
      \midrule

      \multirow{2}{*}{1} 
        & \multirow{1}{*}{MHP}  & 102.9 & 101.3 & 94.3 & 90.7 & 140.8 & 106.1 \\
        & \multirow{1}{*}{LLM}  & 85.3 (+17\%) & 84.9 (+16\%) & 70.9 (+25\%) & 79.3 (+13\%) & 93.3 (+34\%) & 82.7 (+22\%) \\
      \midrule[1.05pt]

      \multirow{4}{*}{2} 
        & \multirow{1}{*}{MHP}  & 73.2 (+29\%) & 65.4 (+35\%) & 66.4 (+30\%) & 56.9 (+37\%) & 95.7 (+32\%) & 71.5 (+33\%) \\
        & \multirow{1}{*}{CoELA}  & 47.3 (+54\%) & \textbf{42.3 (+58\%)} & 50.7 (+46\%) & \textbf{47.0 (+48\%)} & 69.1 (+51\%) & 51.3 (+51\%) \\
        & \multirow{1}{*}{ProAgent}  & \textbf{45.5 (+55\%)} & 54.4 (+46\%) & 46.4 (+51\%) & 54.3 (+40\%) & 69.2 (+51\%) & 53.9 (49\%) \\
        & \multirow{1}{*}{MiTa (ours)}  & 51.1 (+49\%) & 45.4 (+55\%) & \textbf{46.3 (+51\%)} & \textbf{47.0 (+48\%)} & \textbf{54.6 (+61\%)} & \textbf{48.8 (+54\%)} \\
      \midrule[1.05pt]

      \multirow{4}{*}{3} 
        & \multirow{1}{*}{MHP}  & 58.2 (+43\%) & 55.8 (+45\%) & 61.3 (+35\%) & 48.7 (+46\%) & 85.6 (+39\%) & 61.9 (+41\%) \\
        & \multirow{1}{*}{CoELA}  & 40.5 (+60\%) & 34.7 (+65\%) & 35.9 (+62\%) & 37.9 (+58\%) & 47.7 (+66\%) & 39.3 (+63\%) \\
        & \multirow{1}{*}{ProAgent}  & 46.3 (+55\%) & 36.4 (+64\%) & 36.8 (+61\%) & 41.3 (+54\%) & 58.1 (+59\%) & 43.8 (+58\%) \\
        & \multirow{1}{*}{MiTa (ours)}  & \textbf{39.7} (+\textbf{61\%}) & \textbf{30.9} (\textbf{+69\%}) & \textbf{28.4} (\textbf{+70\%}) & \textbf{33.2} (+\textbf{63\%}) & \textbf{39.7} (+\textbf{72\%}) & \textbf{34.4} (+\textbf{68\%}) \\
      \bottomrule[1.2pt]
    \end{tabularx}
    \vspace{-0.4cm}
\end{table*}

\section{Experiment}
\vspace{-0.1cm}
\label{sec:typestyle}

\subsection{Experiment Setup}
\vspace{-0.1cm}

\textbf{Environment.} VirtualHome-Social\cite{watch-and-help} is an interactive 3D simulation platform designed to programmatically model complex household activities. 
The test set comprises five task categories defined in C-WAH, named \textit{Prepare Tea}, \textit{Wash Dishes}, \textit{Prepare a Meal}, \textit{Put Groceries}, and \textit{Set up Table}. We conduct experiments under symbolic and visual observation settings, where the former provides structured object-level information and the latter offers egocentric RGB-D inputs with auxiliary observations. We adopt GPT-4o, Qwen3-Plus and DeepSeek-V3.1 as LLMs in our agents.

\textbf{Baseline.} We adopt three advanced multi-agent frameworks as baselines. \textbf{MHP}\cite{watch-and-help} is an MCTS-based hierarchical planner that performs best in the Watch-And-Help Challenge. \textbf{CoELA}\cite{COELA} is a modular framework for multi-agent cooperation on multi-objective long-horizon tasks. \textbf{ProAgent}\cite{proagent} infers teammates’ intentions from observations and updates its own beliefs based on their actions.

\textbf{Metrics.} We use the average step $L$ required to complete a task as the primary metric for evaluating efficiency. Additionally, we define the Efficiency Improvement ($EI$) as $\Delta M/M_0$, where $\Delta M$ represents the change in the primary efficiency metric, and $M_0$ denotes the greater of the compared metric values to ensure numerical stability.

\vspace{-0.35cm}
\subsection{Main Results and Analysis}
\vspace{-0.15cm}

\textbf{Performance comparison of different frameworks.} We evaluated MiTa with two and three agents, comparing it against single-agent and previous multi-agent frameworks across five tasks (Table \ref{tab 1}). Results show that both MHP and LLM-based single-agent models are significantly less efficient than multi-agent frameworks, highlighting the limitations of single-agent approaches. Compared with CoELA and ProAgent, MiTa achieves a lower average step $L$ and higher efficiency improvement ($EI$), with the advantage most pronounced in the three-agent setting, where $EI$ increases from 63\% (CoELA) and 58\% (ProAgent) to 68\%. 
With two agents, MiTa underperforms in \textit{Wash Dishes} and \textit{Put Groceries}, but achieves shorter average steps than CoELA and ProAgent by 2.5 and 5.1, respectively.
In the three-agent configuration, MiTa attains the best results across all five tasks, highlighting its superior capability in complex, long-horizon scenarios.

\begin{figure}[htbp]
    \centering
    \vspace{-0.1cm}
    \includegraphics[width=0.49\textwidth]{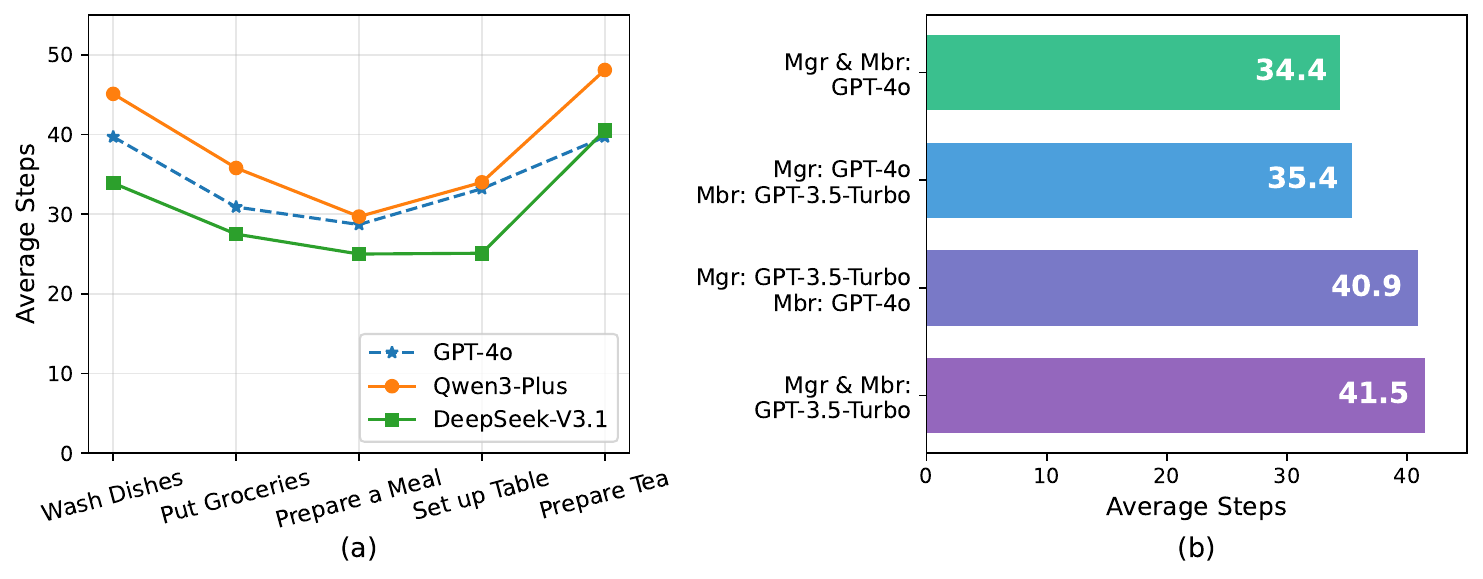} 
    \vspace{-0.7cm}
    \caption{(a) MiTa performance under symbolic observation across different LLMs. (b) Robustness under symbolic observation (3 agents) in computationally limited settings.}
     \vspace{-0.35cm}
    \label{fig.combined_chart}
\end{figure}

\textbf{Effects of different LLMs.}
To assess the robustness of our framework, we replace the backbone LLM for reasoning with GPT-4o, Qwen3-Plus, and DeepSeek-V3.1, keeping all other components fixed.
As shown in Fig.\ref{fig.combined_chart}(a), MiTa achieves stable task execution across all models. Notably, replacing the backbone with more powerful LLMs consistently reduces the required time steps. This suggests that stronger LLMs possess superior planning and coordination capabilities, thereby highlighting the importance of integrating advanced LLMs. 

\textbf{Robustness of resource constraints.}
To further assess the practical applicability of MiTa, we evaluate its robustness under limited-resource settings by assigning lower-capacity LLMs to agents. As shown in Fig.\ref{fig.combined_chart}(b), replacing member agents’ backbone with GPT-3.5-Turbo leads to only a marginal increase of one step in average task completion, indicating that MiTa maintains strong coordination even with weaker reasoning modules. In contrast, when the manager agent is powered by GPT-3.5-Turbo, coordination efficiency drops substantially, by 6.5 and 7.1 steps when paired with members using GPT-4o and GPT-3.5-Turbo, highlighting that the manager relies more heavily on a strong LLM.

\vspace{-0.4cm}
\subsection{Ablation Study}
\vspace{-0.1cm}

To validate the effectiveness of each component in the MiTa framework, we conducted ablation studies, where each variant excludes a specific module from the full system. As shown in Table \ref{tab:ablation}, MiTa w/o allocation performs the worst, with time steps increased by 14 and 4.7 under symbolic and visual observations, respectively. This degradation highlights the importance of centralized coordination provided by the manager, as agents without task allocation tend to act myopically, prioritizing their own local states rather than selecting globally optimal actions. Compared with MiTa, the w/o summary variant results in efficiency drops of 15.7\% and 3.2\% under symbolic and visual observations, confirming the effectiveness of memory integration.

%

\begin{table}[htbp]
  \centering
  \vspace{-0.45cm}
  \caption{Average steps with different component combinations under symbolic and visual observations.}
  \vspace{0.1cm}
  \label{tab:ablation}
      \begin{tabular}{l c c}
        \toprule[1.15pt] 
        \textbf{Methods} & \textbf{Symbolic Obs($\downarrow$)} & \textbf{Visual Obs($\downarrow$)} \\
        \midrule
        MiTa (ours)     & \textbf{34.4}  & \textbf{68.24} \\
        w/o allocation  & 48.4  & 72.94   \\
        w/o summary    & 39.8  & 70.4  \\
        \bottomrule[1.15pt] 
      \end{tabular}
\vspace{-0.4cm}
\end{table}


\vspace{-0.2cm}
\section{CONCLUSIONS}
\vspace{-0.2cm}

\label{sec:refs}
In this work, we propose MiTa, a Memory-integrated Task allocative multi-agent framework. By combining negotiation-aware allocation with an episodic summary module, MiTa balances local insights with global consistency and preserves long-term contextual information. Experimental results show that MiTa achieves shorter steps and more effective coordination over existing methods, while delivering robust results even under weaker member models, highlighting its potential for deployment in computationally limited scenarios.

\bibliographystyle{IEEEbib}
\bibliography{strings,refs}



\end{document}